\documentclass[runningheads]{llncs}
\usepackage{todonotes}
\usepackage{tikz}
\usepackage{tikz-uml}
\usepackage{adjustbox}
\usepackage{float}
\usepackage{caption}
\usepackage{subcaption}
\usepackage{graphicx}
\usepackage{subfloat}
\usepackage{float}
\usepackage{multirow}
\usepackage{soul}
\usepackage[T1]{fontenc}
\usepackage{graphicx}

\begin{document}
\title{Storage Management with Multi-Version Partitioned BTrees}
\subtitle{[Extended Version]}

\titlerunning{Storage Management with MV-PBT [Extended Version]}

\author{Christian Riegger \and
Ilia Petrov}
\authorrunning{C. Riegger \and I. Petrov}
\institute{Reutlingen University
\email{\{christian.riegger,ilia.petrov\}@reutlingen-university.de}\\
\url{https://www.dblab.reutlingen-university.de}
}
\maketitle              % typeset the header of the contribution
\begin{abstract}
Database Management Systems and K/V-Stores operate on updatable datasets -- massively exceeding the size of available main memory. Tree-based K/V storage management structures became particularly popular in storage engines. B$^+$-Trees \cite{btree,bplustree} allow constant search performance, however write-heavy workloads yield in inefficient write patterns to secondary storage devices and poor performance characteristics. LSM-Trees \cite{lsm_sears,lsm_oneil} overcome this issue by horizontal partitioning fractions of data -- small enough to fully reside in main memory, but require frequent maintenance to sustain search performance. 
Firstly, we propose Multi-Version Partitioned BTrees (MV-PBT) as sole storage and index management structure in key-sorted storage engines like K/V-Stores. Secondly, we compare MV-PBT against LSM-Trees. The logical horizontal partitioning in MV-PBT allows leveraging recent advances in modern B$^+$-Tree techniques in a small transparent and memory resident portion of the structure. Structural properties sustain steady read performance, yielding efficient write patterns and reducing write amplification.

We integrated MV-PBT in the WiredTiger \cite{wiredtiger_mdb} KV storage engine. MV-PBT offers an up to 2x increased steady throughput in comparison to LSM-Trees and several orders of magnitude in comparison to B$^+$-Trees in a YCSB \cite{ycsb} workload.

\keywords{Storage Engine \and Storage Management \and Append Storage.}
\end{abstract}
\section{Introduction}

High performance persistent key-sorted No-SQL storage engines became the load-bearing backbone of online data-intensive applications. Such engines exist as standalone K/V-Stores (Key/Value Stores) \cite{rocksdb,wiredtiger_mdb} as well as in integrated in DBMS storage engines \cite{bwtree,mongodb,myrocks}. Obviously, backing tree-based K/V storage management structures -- i.e. B$^+$-Trees \cite{btree}, LSM \cite{lsm_oneil,lsm_sears} and derivatives \cite{prefix_bayer,bwtree} --  natively enable necessary advanced lookup operations beside equality search, e.g. key prefix or inclusive and exclusive range searches, with (nearly) constant logarithmically scaling performance characteristics. Continuous modifications require special care to preserve constant performance characteristics and mentioned search features. Although B$^+$-Trees offer constant search performance to data in main memory and on secondary storage devices, modifications yield in inelastic performance characteristics. LSM-Trees sacrifice properties of a single tree structure to overcome this issue by buffering modifications in a fraction of main memory, typically tree-based components, and leveraging flash-based secondary storage device characteristics on eviction and necessary background merge operations. 

\subsubsection{Flash technology in SSD secondary storage devices} exhibit individual characteristics. I/O operations possibly are independent or decomposed executed in multiple structural levels of an SSD, whereas a high internal parallelism and I/O-performance is enabled \cite{flash_parallelism3,flash_parallelism_die1,flash_parallelism_plane}. However, reads perform an order of magnitude better than writes, yielding in a asymmetric I/O behavior. Whilst reads perform nearly identical for random and sequential access patterns, write I/O is preferably sequentially performed  \cite{flash_characteristics}. Furthermore, pages are replaced out-of-place, wherefore much slower erases and background garbage collection is necessary \cite{flash_asymmetric,sias_diss}.  

\subsubsection{B$^+$-Trees and derivatives} achieve a constant logarithmically scalable search performance, since root-to-leaf traversal operations depend on their height -- even in case of massive amounts of stored data records. Commonly used inner nodes of traversal paths allow fast access to data in leaf nodes with few successive read I/O. However, B$^+$-Trees are probably vulnerable in case of modifications. Whilst insertions, updates and deletions of records possibly facilitate steady throughput in main memory by optimized and highly scalable maintenance procedures \cite{bwtree,wiredtiger_mdb}, massive amounts of maintainable key-sorted data yield in random write I/O and high write amplification on secondary storage devices once modifications get persisted on eviction of '\textit{dirty}' buffers. In order to preserve strict lexicographical sort order of records, maintenance operations cause cascading node splits, whereby blank space is created to accommodate additional separator keys in inner nodes and records in leaves in the designated arrangement. As a result, sub-optimally filled nodes reduce cache efficiency and contained information is written multiple times, yielding in a high write and space amplification. Furthermore, read I/O on secondary storage devices of partially filled nodes lead to high read amplification. Therefore, for massive amounts of contained data, B$^+$-Trees become write-intensive, even in case of proportionately few modifications, yielding in following \textbf{problems}:
\begin{itemize}
	\item low benefit from main memory optimizations, since nodes are frequently evicted
	\item low cache efficiency and high read amplification due to partially filled nodes
	\item massive space and write amplification on secondary storage devices
\end{itemize}

\subsubsection{Alternatively, LSM-Trees are optimized for high update rates} and obtain a sequential write pattern, since modifications are buffered in tree-based LSM components in main memory. Components get frequently switched, merged and evicted to persistent secondary storage devices. Generally, background merge operations counteract the data fragmentation and increased read and search effort, however this behavior also increases its write amplification. Several approaches in merge policies \cite{lsm_sears} and reduction of read amplification \cite{surf,rosetta,crizz_bloomrf_old,remix} have been introduced. Certainly, flash allows high internal parallelism and multiple reads of parallel traversal operations. Nevertheless, since components are separate structures, they effectively leverage neither caching effects on traversal nor logarithmic capacity capabilities per height of B$^+$-Trees. Moreover, creation of new components on switch procedure is not transparent to the storage engine and relies on high-level maintenance of the database schema. Finally, due to append-based record replacement technique in LSM, key uniqueness is assumed, wherefore the application in storage engines of DBMS with non-unique indexes is complicated. \textbf{Challenges in LSM} are defined as follows:

\begin{itemize}
	\item inefficient caching behavior of decoupled components require frequent merges and yield in considerable write amplification
	\item hence, high internal parallelism of flash is not leveraged for read operations
	\item components are non-transparent for further layers of a storage engine
	\item non-unique indexing requires additional care
\end{itemize}

\subsubsection{We propose Multi-Version Partitioned BTree (MV-PBT)} as sole storage and index management structure in KV-storage engines. MV-PBT is based on Partitioned BTrees (PBT) \cite{pbt_sorting_indexing_graefe}, an enhancement of a traditional B$^+$-Tree. (MV-) PBT relies on manipulation of an artificial leading key column of every record -- the partition number; and exploiting the regular lexicographical structure of B$^+$-Trees for partition management. Recent publications introduced (MV-) PBT as a highly scalable indexing structure in DBMS with multi-version concurrency control (MVCC) and massive index update pressure \cite{crizz_PBT_write,crizz_athen,crizz_htap}. However, this paper focus on MV-PBT as sole storage management structure in KV-storage engines. The contributions are:
\begin{itemize}
	\item Diminishing write amplification in append-based storage management with MV-PBT by sequential write of saturated partition managed nodes
	\item Transparent internal partition management and atomic partition switch operations without schema maintenance requirements
	\item Single root node as entry point in the B$^+$-Tree structure allows to leverage logarithmic capacity and commonly cached and traversed inner nodes
	\item Reduction of merge-triggered write amplification and accompanying pressure on secondary storage devices by Cached Partitions
	\item Leveraging scalable in-memory optimizations and compression techniques of B$^+$-Tree structures for massive amounts of data in a very hot fraction
	\item Prototypical implementation and experimental evaluation in WiredTiger \cite{wiredtiger_mdb}, which provides competitive B$^+$-Tree and LSM-Tree implementations
\end{itemize}

\subsubsection{Outline}
We present an architectural overview of MV-PBT in Section \ref{sec:mvpbt_arch}. Sections \ref{sec:cp} and \ref{sec:gc} focus on reduction of write amplification by data skipping and fast retrieval in a horizontally partitioned structure and considering defragmentation only as a result of garbage collection. We evaluated the storage management structures in the homogeneous storage engine WiredTiger 10.0.1 in Section \ref{sec:eval} and conclude in Section \ref{sec:conclude}.

\section{Architecture of Multi-Version Partitioned BTrees}
\label{sec:mvpbt_arch}

\begin{figure}[t]
	\centering
	\adjustbox{trim={.097\width} {.36\height} {.095\width} {.02\height},clip,width=.85\textwidth}
	{\includegraphics{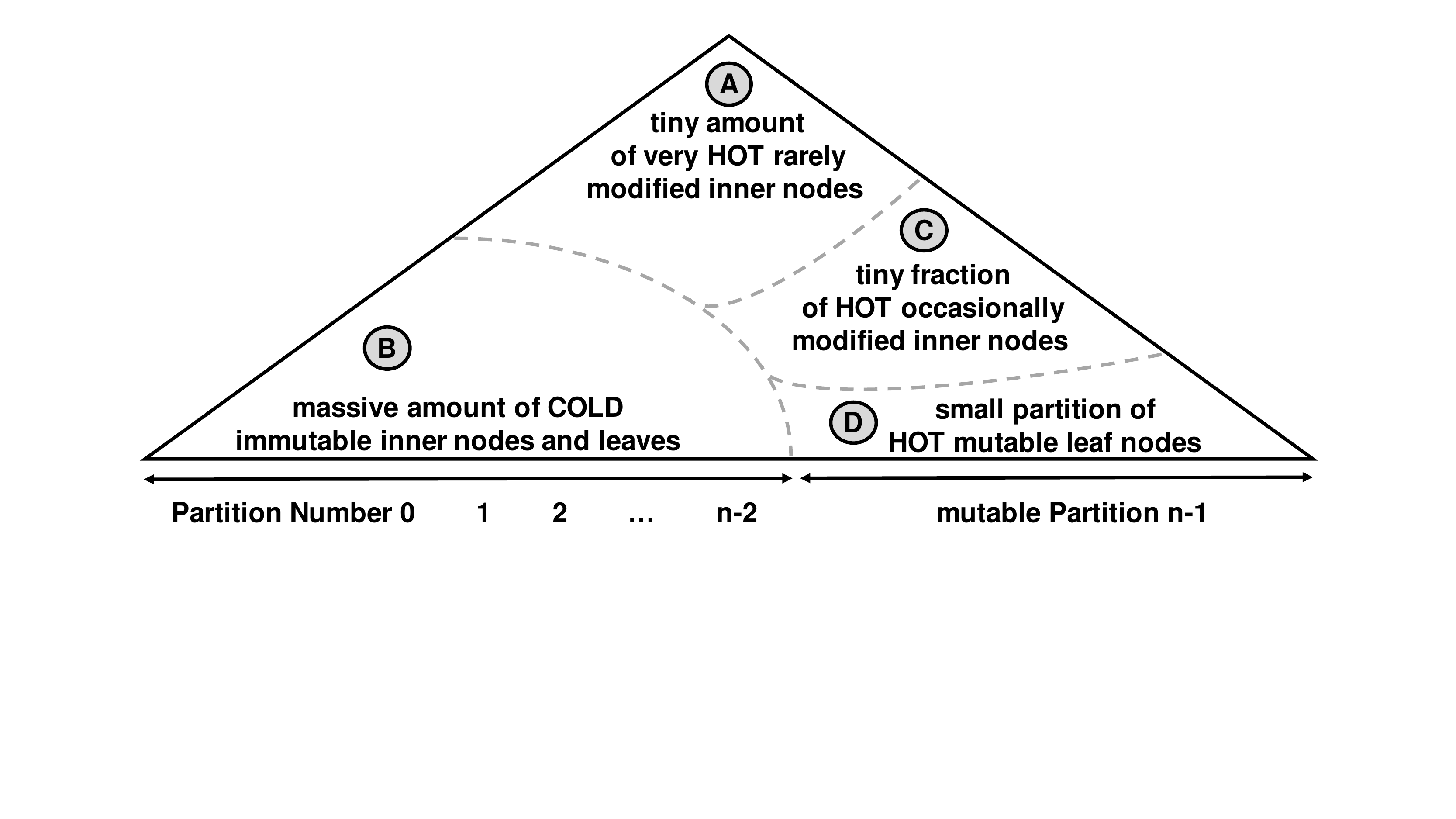}}
	\caption{Logical horizontal partitioning in MV-PBT and Replacement Policy of MV-PBT-Buffer yield in hot/cold separation within one single tree structure and ultimately enables a sequential write pattern of whole partitions.}
	\label{fig:pbt_hot_cold}
\end{figure}

Multi-Version Partitioned BTree (MV-PBT) as an append-based and version-aware storage and indexing structure relies on well-studied algorithms and structures of traditional B$^+$-Trees -- with which they share many characteristics and areas of application. Therefore, MV-PBT is able to adopt and even leverage characteristics of advances in modern B$^+$-Tree techniques. The proposed approach facilitates straightforward horizontal partition management within one single B$^+$-Tree structure in order to keep a very hot mutable fraction of leaves in fast volatile main memory (compare Fig. \ref{fig:pbt_hot_cold}) -- the MV-PBT-Buffer including the most recent partition leaves is temporarily apart from the regular buffer replacement policy. Reaching a certain dirty memory footprint threshold initiates an atomic partition switch operation, which asynchronously finalizes in a sequential write of dense-packed cleaned data in leaves and referring inner nodes, in order to interference-freely absorb ongoing modifications. Since partitions are principally defined by the existence of associated records, they appear and vanish as simply as inserting or deleting records \cite{pbt_sorting_indexing_graefe}, however, auxiliary meta data structures allow a massive speed-up of operations. Append-based structures allow modifications of already persisted data by out-of-place replacement. MV-PBT enhances this behavior by additional record types, which allow internal indexing and non-uniqueness of data and enables native B$^+$-Tree-like indexing features. Moreover, maintenance of multiple record circumstances imply the adoption of multi-version capabilities by the assignment of transaction timestamps in MVCC with snapshot isolation. Low write amplification, sequential writes of dense-packed nodes, commonly utilized inner nodes with one single root as entry point, parallelized multi-partition search operations as well as multi-version indexing capabilities make MV-PBT superior as sole storage and index management structure in storage engines.

\subsubsection{MV-PBTs Auxiliary Data Structures} 
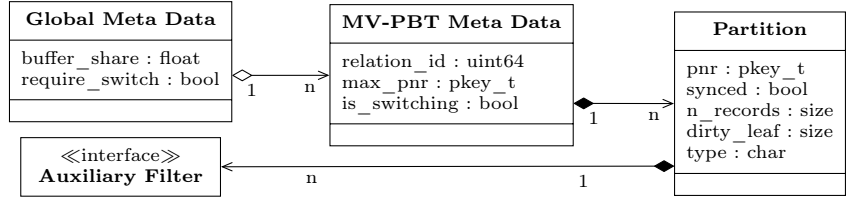
\begin{figure}[t]
	\tikzumlset{fill class=white, font=\scriptsize}
	\begin{center}
		\begin{tikzpicture}
			\umlclass[width=19.5ex]{Global Meta Data}{
				buffer\_share : float \\
				require\_switch : bool
			}{}
			\umlclass[x=4.4,y=-.18]{MV-PBT Meta Data}{
				relation\_id : uint64 \\
				max\_pnr : pkey\_t \\
				is\_switching : bool
				
			}{}
			\umlclass[x=8.5,y=-.559]{Partition}{
				pnr : pkey\_t \\
				synced : bool \\
				n\_records : size \\
				dirty\_leaf : size \\
				type : char \\
			}{}
			\umlsimpleclass[width=19.5ex,x=0,y=-1.4,type=interface]{Auxiliary Filter}
			\umluniaggreg[geometry=--,anchor1=-7,mult1=1,mult2=n]{Global Meta Data}{MV-PBT Meta Data}
			\umlunicompo[geometry=--,anchor1=-13.8,mult1=1,mult2=n]{MV-PBT Meta Data}{Partition}
			\umlunicompo[geometry=--,anchor1=-144.5,arg1=1,arg2=n]{Partition}{Auxiliary Filter}
			
		\end{tikzpicture}
	\end{center}
\vspace{-0.4cm}
	\caption{Auxiliary recoverable MV-PBT data structures.}
	\label{fig:pbt_cached_structures}
\end{figure}

information is entirely contained in the B$^+$-Tree structure. For instance, the mutable most recent partition number could be identified by searching the rightmost record in the tree structure. Since cached information is frequently required and its memory footprint is very low, auxiliary data structures are cached in RAM (an excerpt is depicted in Fig. \ref{fig:pbt_cached_structures}). MV-PBT data structures require neither locking for any atomic operation nor additional logging of modifications, since the lightweight information is completely recoverable from basic B$^+$-Tree by a scan operation. All information of horizontal partitioning is anchored within the tree structure, i.e. horizontal partitioning is transparent to further storage engine modules -- contrary to schema modifications in LSM-Trees.

Multiple MV-PBT exist within a storage engine, which commonly share the MV-PBT-Buffer threshold. The MV-PBT Meta Data belongs to a specific relation in the schema. Its most recent partition number (\texttt{max\_pnr}) is frequently required to determine record prefixes as well as for atomic switching operation. An MV-PBT comprises of several valid partitions, which contain a set of meta data like the number of records or specific partition type characteristics. Finally, auxiliary filter structures for point and / or range queries are referenced; e.g. fence keys, (prefix) bloom filters or hybrid point and range filters \cite{surf,rosetta,crizz_bloomrf_old}. 

\subsubsection{Partition Number Prefixes} 
\begin{figure}[t]
	\begin{minipage}{.5\linewidth}
		\centering
			\subfloat[Record with Partitioned Key]{\vspace{-0px}\label{fig:pkey_artificial_horizontal_partitioning}
			\adjustbox{trim={.0\width} {.33\height} {.0\width} {.0\height},clip,width=\textwidth}
			{\includegraphics{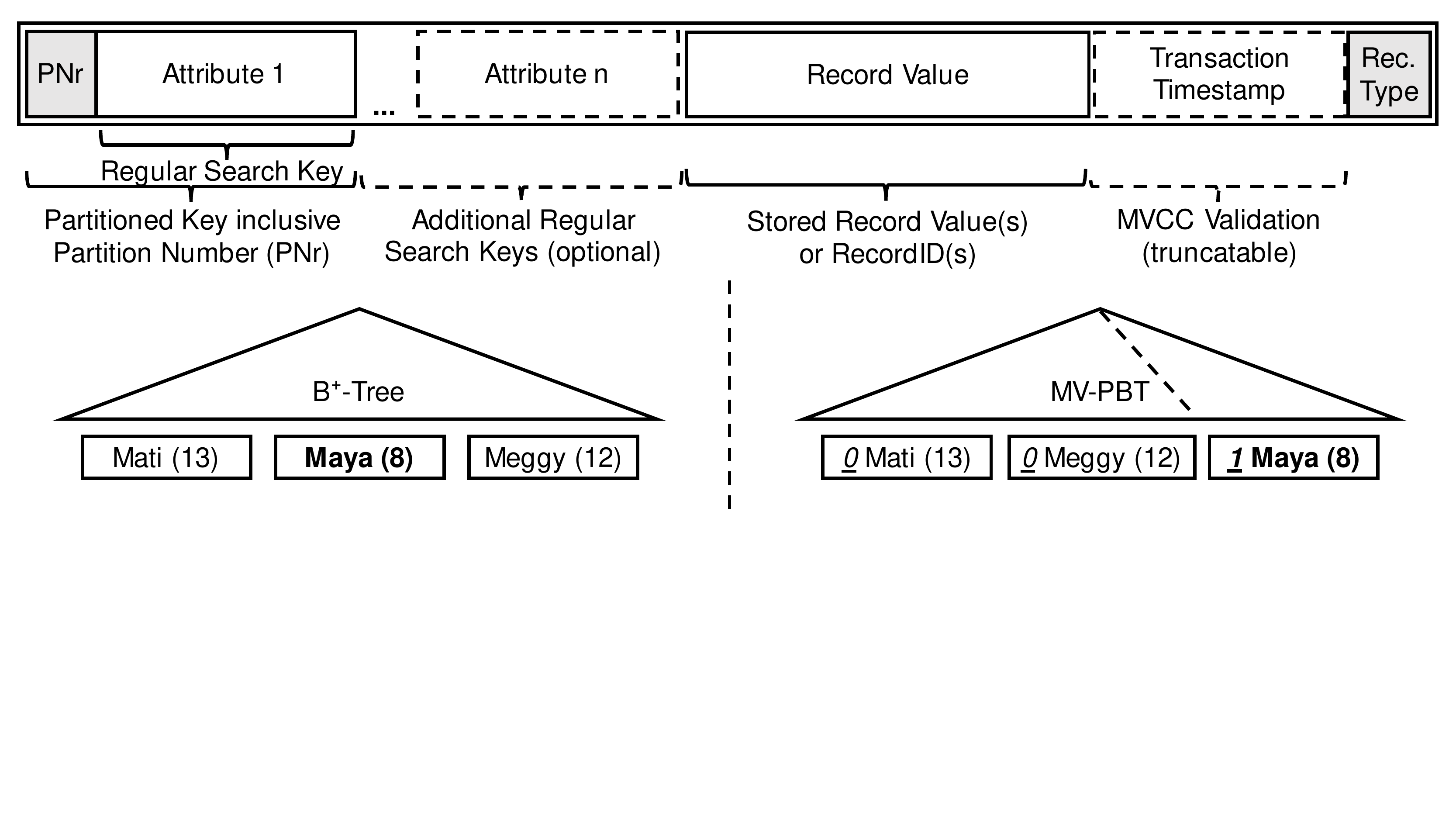}}}
	\end{minipage}
	\begin{minipage}{.5\linewidth}
		\centering
		\subfloat[Comparison Costs for both approaches]{\vspace{-0px}\label{fig:pkey_artitficial_cost}
			\adjustbox{trim={.0\width} {.33\height} {.0\width} {.0\height},clip,width=\textwidth}
				{\includegraphics{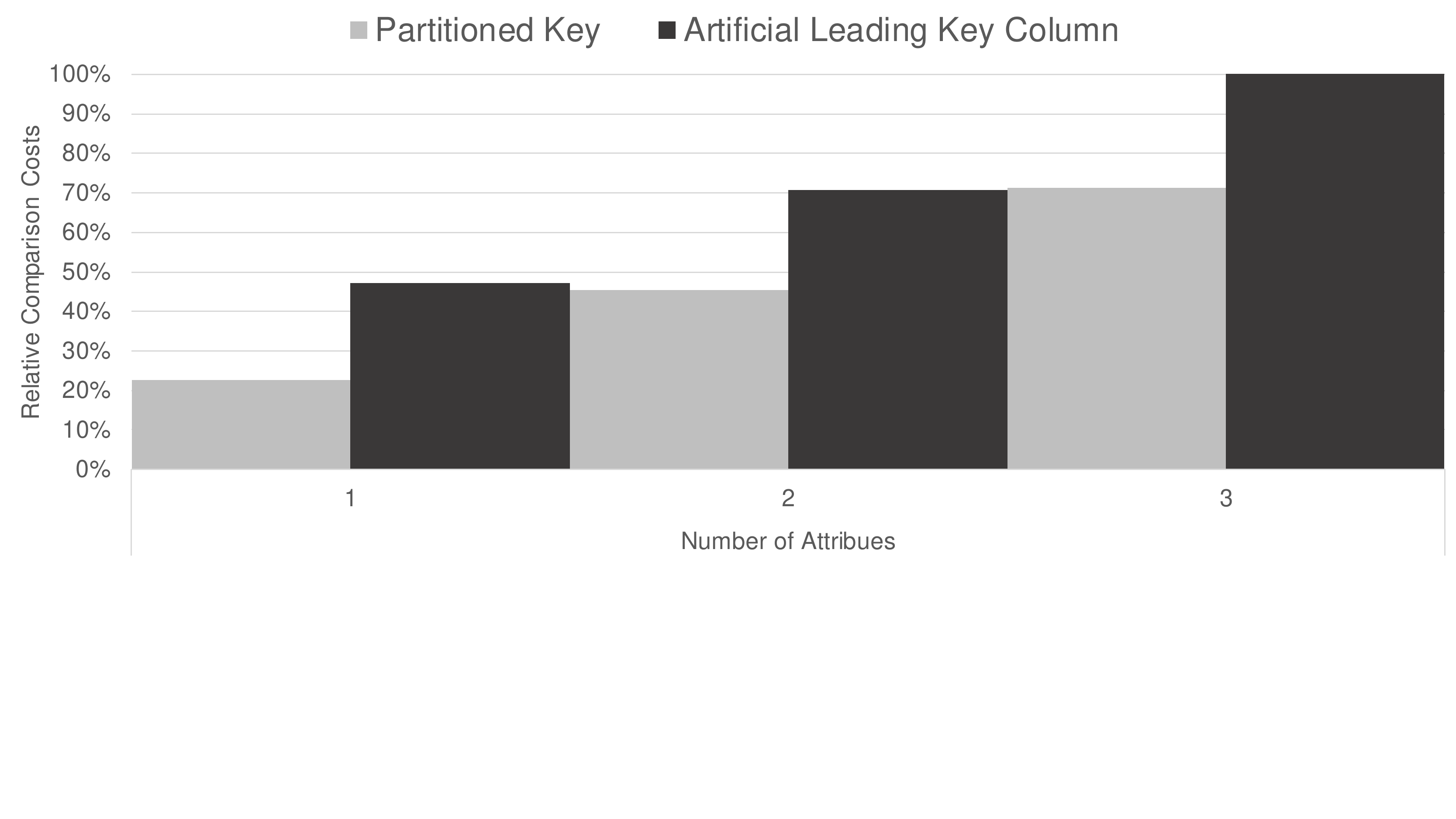}}}
\end{minipage}\par\medskip
\vspace{-0.25cm}
\caption{Horizontal partition maintenance with Partitioned Keys}
	\label{fig:pkey_artificial}
\end{figure}

are prepended to each record key with the central scope of leveraging lexicographical sort capabilities of B$^+$-Trees in order to achieve a logical horizontal partitioning. Partition numbers could be of any comparable data type, e.g. 2 or 4-byte integers, and might are maintained in an artificial leading key column \cite{pbt_sorting_indexing_graefe}. However, combining the partition number and the first record key attribute in a \textit{partitioned key} type (compare Fig. \ref{fig:pkey_artificial}.a) enables cache efficient comparison of co-aligned attributes as evaluated in Fig. \ref{fig:pkey_artificial}.b. Additional storage costs are negligible due to prefix truncation techniques. \textit{Partitioned keys} are simply allocated when setting search keys and their prefix becomes hidden by returning an offset in the leading key attribute in order to retain transparent horizontal partitioning.

\subsubsection{Multi-Version Capabilities} accompanying well the out-of-place replacement in MV-PBT. Multi-Version Concurrency Control (MVCC) with Snapshot Isolation (SI) are a common technique to enable high transactional parallelism in storage engines, since readers and writers are not mutually blocking as each transaction operates on a separate snapshot of data. Therefore, multiple versions records of one logical tuple are maintained in a version chain -- each is valid for a different period in time. MV-PBT adopts a new-to-old ordering approach of physically materialized version records with out-of-place update scheme and one-point invalidation model \cite{crizz_htap,sias_chains} -- i.e. predecessor versions remain unchanged on modification, whereas write amplification is massively reduced. Successor version records are annotated with the current transaction timestamp (which may become truncated on eviction to secondary storage devices, whenever no preceding snapshot is active) and are inserted in the most recent partition in the MV-PBT-Buffer. Thereby, it is possible to maintain multiple version records in one partition, e.g. as separate record \cite{crizz_htap} or in-memory update lists \cite{wiredtiger_mdb}. Based on the logical search succession in MV-PBT from new-to-old, transaction snapshots identify their visible version record and skip others, based on the annotated transaction timestamps. Since record data values are physically materialized in each version record, identified records are directly applicable. 

\subsubsection{Record Types in MV-PBT} feature all operations over logical tuple life-cycle without modifying predecessor version records. During lifetime, it gets created, modified and deleted while it is frequently read. \textit{Regular Records} declare the begin of the life-cycle, hence there is no predecessor version. Its transaction timestamp is applied by the inserting transaction and indicates its validation. \textit{Replacement Records} indicate a new record value on update. Its timestamp invalidates its predecessor as well as validates itself. Replacement Records are also applied on modifications to the record key, however, invalidation requires an \textit{Anti Record} with the predecessor key attribute values and the current transaction timestamp for invalidation. Replacement Records as well as Anti Records probably store its predecessor value for logical tuple assignment as needed in non-uniqueness index management constraints, however, modifications to the key attributes and non-uniqueness indexing constraints with index-only visibility checks \cite{crizz_htap} allow MV-PBT to serve as sole storage and index management structure in storage engines but is out of scope in this paper. Finally, \textit{Tombstone Records} are inserted on deletion of a logical tuple. Major difference to Anti Records is, that successor version records are impossible.

\subsubsection{Atomic Partition Switch and sequential write} of dense-packed leaves and referring inner nodes bring a leading edge in MV-PBT. The whole procedure consists of several partially parallelizable stages. After \textit{(a) determination of switch requirement} by a certain dirty buffer threshold in the MV-PBT-Buffer, a \textit{(b) valuable MV-PBT victim partition} is selected for eviction. Contrary to LSM-Trees, MV-PBT partitions become immutable and switched by \textit{(c) atomically incrementing the most recent partition number} (\texttt{max\_pnr}) in the meta data, since the required B$^+$-Tree structure is already existent and logged anyways. 

\begin{figure} [t]
	\adjustbox{trim={.0\width} {.81\height} {.0\width} {.0\height},clip,width=\textwidth}
	{\includegraphics{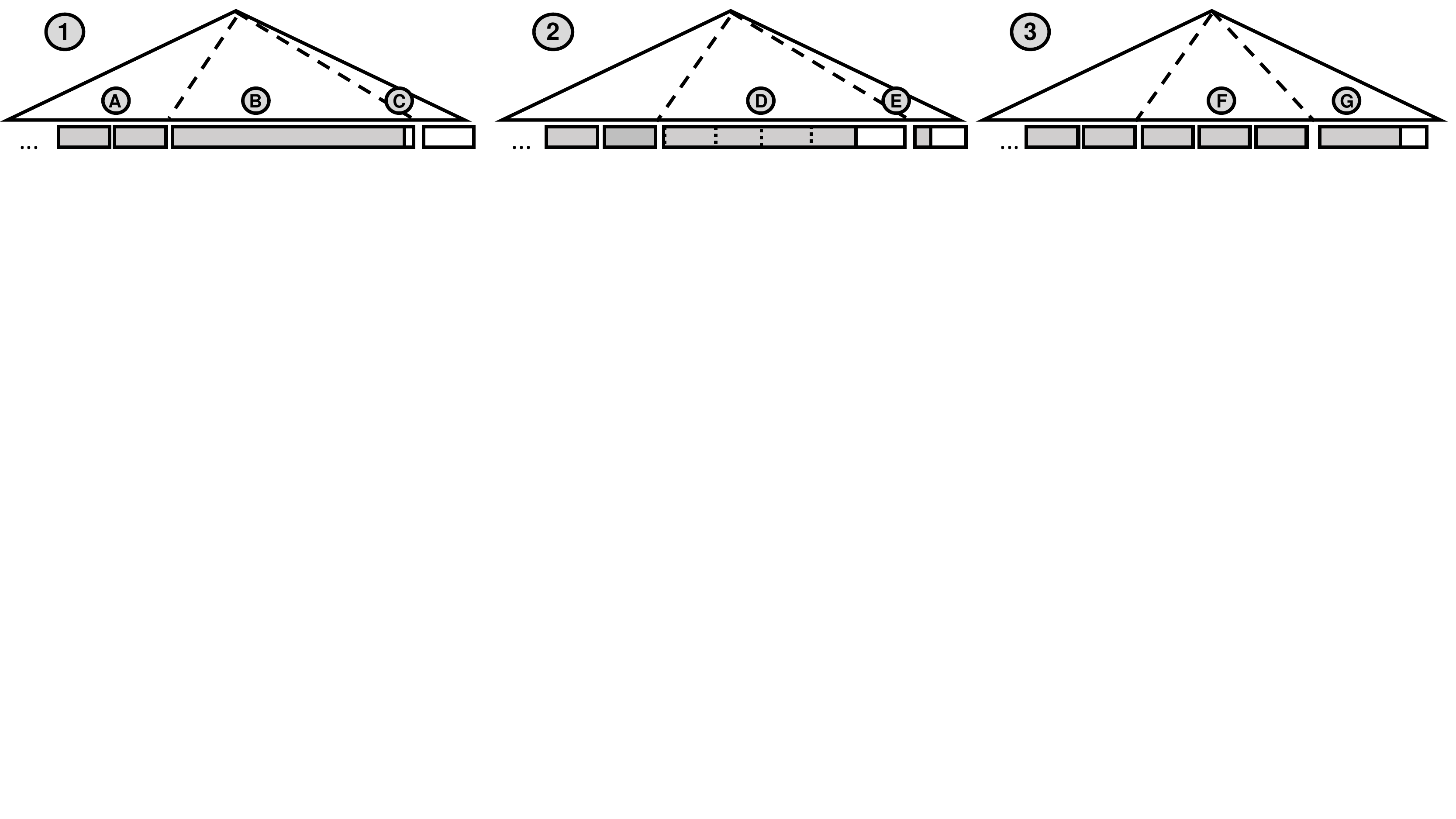}}
	\caption{(1) After atomic partition switch, an MV-PBT consists of (A) persistent, (B) a victim and (c) a most recent partition. Internal nodes and leaves of the victim partition delay maintenance effort (e.g. split operations) by flexible page size until a reconciliation process (2.D). The (E) most recent partition consumes ongoing modifications. Finally, (3) the (F) victim partition is sequentially written to secondary storage and (G) is the only memory mapped partition.}
	\label{fig:pbt_dense_wt}
\end{figure}

\begin{figure}[b]
	\adjustbox{trim={.1\width} {.77\height} {.08\width} {.0\height},clip,width=\textwidth}
	{\includegraphics{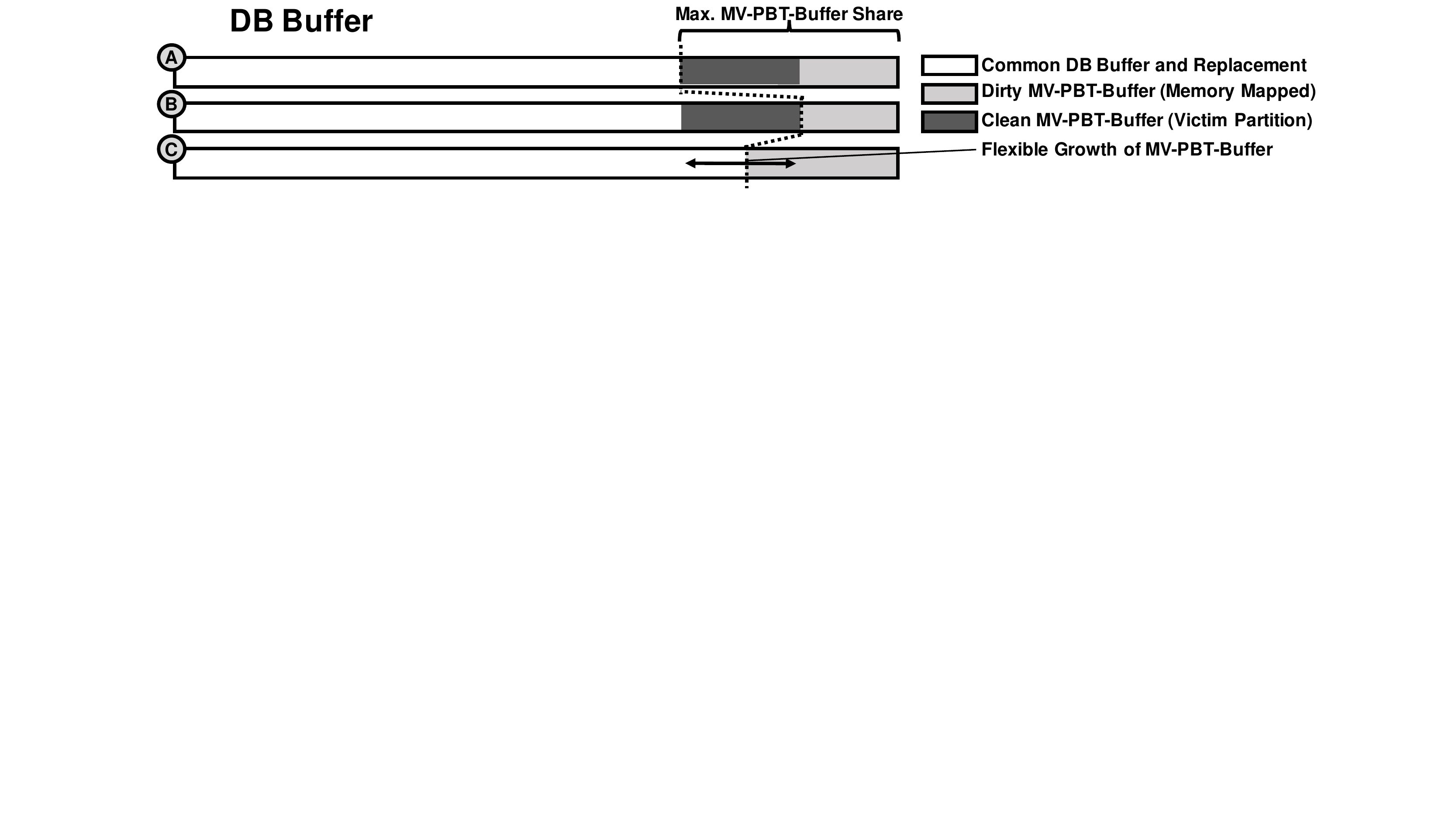}}
	\caption{Flexible MV-PBT-Buffer Share allows cache preserving handover of a clean Victim Partition from (A) the MV-PBT-Buffer to (B) a common buffer replacement policy and (C) flexible growth up to a max. MV-PBT-Buffer Share.}
	\label{fig:pbt_flexible_growth}
\end{figure}

However, records are probably not yet in their final \textit{(d) defragmented and dense-packed disk layout}, since structure modifications are the result of a randomly inserting workload. One approach to avoid expensive partition-internal structure modifications (e.g. node merges) is to simply re-inserting the still valid contents in their final arrangement by manipulating the partition number in a bulk load operation \cite{crizz_htap}. B$^+$-Trees allow efficient split policies to support high fill factors by this operation. Finally, visibility characteristics of both partitions are swapped and the randomly grown source partition gets cropped from the tree. Another approach is to leverage modern B$^+$-Tree techniques. In order to avoid structure modifications, referenced main memory nodes are allowed to flexibly grow and finally get divided and structured in the disk layout in a reconciliation process (depicted in Fig. \ref{fig:pbt_dense_wt}).

Auxiliary \textit{(e) filter structures} are generated as a natural by-product of defragmentation and dense-packing, since records are accessed anyways. Whenever (a fraction of) leaf nodes obtained their final layout, it is possible to \textit{(f) perform a sequential write of leaves and referring inner nodes} by traversing the tree structure and following the sibling pointers -- yielding in a bottom-up sequential write of nodes, level by level. Finally, the persisted leaves are \textit{(g) passed to the regular replacement policy} in order to sustain a constant buffer factor and memory footprint (Fig. \ref{fig:pbt_flexible_growth}).

\subsubsection{Basic Operations} in MV-PBT are based on regular a B$^+$-Tree -- i.e. they have logarithmic complexity. Every modifying operation is treated as an insertion of a record of a respective type. Thereby, the current transaction timestamp is set for validation in visibility checks -- and one-point-invalidation of conceivable predecessors, respectively, which can be located in a preceding or the current partition. However, due to the partitioned key, each modifying operation is performed in the most recent partition in main memory. This is also valid in case of concurrent partition switch by overwriting the partition number of an insertion record key and immediate re-traversal from root. Additional constraint support is very uncommon in pure storage management since records are typically overwritten by blind insertions, however, this is facilitated by MV-PBT in preceding equality search operations.

Equality and range search operations perform root-to-leaf traversals of a (sub-)set of partitions by manipulation of the partition number in the partitioned search key. Partitions are preselected by auxiliary filter structures. Logically, partitions are searched in reverse order from the most recent to the lowest numbered one. Based on the selectivity of a query, partitions may are sequentially processed or by parallel traversals in a merge sort operation. In case of equality searches, sequential processing allow minimal read amplification, contrary, sorted range searches favorably adopt the merge sort approach, whereby  multiple cursors are applied and get individually moved and returned to a higher level merge sort cursor. Thereby, record transaction timestamps are checked for visibility to a transaction snapshot. Based on a regular visibility check, invisible  and invalidated records are skipped, invalidating records are remembered for exclusion of occurring predecessors (which are subsequently accessed) and matching records are returned \cite{crizz_htap}.  

\section{Cached Partition: Stop Re-Writing valid Data}
\label{sec:cp}
MV-PBT introduces a logical horizontal partitioning within one single tree structure in order to leverage characteristics of secondary storage devices. This data fragmentation influences the search operations in different ways. Obviously, several possible storage locations of a requested record implies additional search effort. Actually regular B$^+$-Trees incur increased search costs in randomly grown structures, due to diminishing cache efficiency of partially filled inner nodes. Contrary, LSM-Trees keep a read-optimized layout within each component, however, multiple entry points and referenced inner nodes are neither commonly cached nor leverage logarithmic capacity. LSM-Trees counteracting increased search effort with background merge operations, whereby write amplification of still valid data is increased. 

MV-PBT preserves a read-optimized and cache-efficient layout for immutable nodes (Fig. \ref{fig:pbt_hot_cold}.B) with one commonly shared entry point and referenced inner nodes (Fig. \ref{fig:pbt_hot_cold}.A) which are subjecting to a optimal fill factor, since append based behavior of referenced data allows efficient split policies (equal to bulk loads). As outlined in Section \ref{sec:mvpbt_arch} (\textit{Atomic Partition Switch and sequential write}), mutable inner nodes and leaves (Fig. \ref{fig:pbt_hot_cold}.C and \ref{fig:pbt_hot_cold}.D) are a hot fraction which sustains maintenance operations of the random workload, however, modern B$^+$-Tree techniques allow main memory efficient delay of maintenance operations. Since the small fraction of inner nodes is commonly used, they are well cached, so that a large portion of the parallel traversal operations is performed without read latencies from secondary storage devices. Successive read I/O in multiple partitions leverage parallelism in flash persistent storage. Moreover, search performance in MV-PBT relies on data skipping by auxiliary filter structures. As a combined result, MV-PBT is able to sustain comparable search performance for higher fragmentation as in LSM-Trees. 

However, variety of auxiliary filter structures imply caching and probe costs as well as massive amount of traversal operations result in high read I/O costs and shrink performance due to growing fragmentation. Instead of adversely re-writing still valid data records in a consolidated arrangement, due to asymmetry of flash and write amplification, MV-PBT introduces \textit{Cached Partitions}. They are an internal index partition, whose records reference a preceding partition, containing the latest version record of a logical tuple in a lexicographical sort order. Several Cached Partitions may exist for a different subset of small partitions and are cyclically created while the MV-PBT evolves. Cached Partitions are the result of a background merge sort of contents in several immutable lower numbered partitions with the respective partition number as value or the contents of several preceding Cached Partitions. Background merge sort results are bulk inserted in an '\textit{invisible}' partition while proceeding, can be paused and continue without wasting work and become finally visible by an atomic status switch. 

Since a subset of partitions is fully indexed in a Cached Partition, a subsequent search operation is able to traverse the subset on the commonly cached path as needed, based on the results of the internal partition index. Cached Partitions assume responsibilities of auxiliary filter structures and allow to exclude the subset of indexed partitions from the regular logical search succession, whereby comparison costs in an internal merge sort are reduced -- the effort is focused on non-indexed and Cached Partitions. Furthermore, cached index records are very space and cache efficient in the search process, since they consist of the key and one partition number (e.g. 2 or 4-byte integer) in a dense-packed arrangement. 

\section{Garbage Collection and Space Reclamation}
\label{sec:gc}
Datasets and tuple values evolve over time. Storage management structures with out-of-place update approaches allow beneficial sequential write patterns and low write amplification, however, invalidated predecessor record versions remain existent on update. Search operations are able to exclude invalid version records from the result set, though visibility checking entail additional processing. Furthermore, version records which are not visible to any active transaction snapshot entail space amplification and additional storage costs. 

In MV-PBT, additional search costs due to fragmentation by horizontal partitioning is well covered by Cached Partitions for insertion of new tuple version records. However, modifications to logical tuple values leave persisted obsolete version records behind, yielding in space amplification. Ideally, obsolete version records are discarded as part of the dense-packing phase on partition switch, however, many version records  become invalidated after they were persisted. For the only reason of space reclamation, MV-PBT occasionally performs a garbage collection (GC) process. Similar to the creation of a Cached Partition, GC is performed by a background merge sort and bulk load operation in a not yet visible partition. Certainly, the stored record value is the regular value of the most recent record version of a tuple. As well, the GC process can throttle and continue without wasting work, since the partition is not yet accessible for querying. After the successful completion, the partition becomes visible and the records of purified preceding partitions become invalidated. Once every active search operation finished, the purified partitions are cropped from the tree structure by an efficient range truncation \cite{wiredtiger_mdb}. 

\section{Experimental Evaluation}
\label{sec:eval}
We present the analysis of MV-PBT as storage management structure in comparison beside the baselines LSM-Trees and B$^+$-Trees fully integrated in WiredTiger 10.0.1 (WT) \cite{wiredtiger_mdb}. LSM-Trees in WT build upon components of the provided B$^+$-Trees upon which MV-PBT is also implemented. A good comparability is achieved, since all structures commonly operate on equal code lines and B$^+$-Tree techniques, e.g.: prefix truncation, suffix truncation and snappy compression; reduced maintenance effort due to flexible page sizes; main memory page representation with sorted areas, update-lists and insertion skiplists; MVCC transaction timestamps in main memory record representation; tree-based buffer management.

\subsubsection{Experimental Setup.} We deployed WiredTiger(WT) 10.0.1 and WT with MV-PBT as storage management structure on an \textit{Ubuntu 16.04.4 LTS} server with an eight core \textit{Intel(R) Xeon(R) E5-1620} CPU, 2GB RAM and an \textit{Intel DC P3600} enterprise SSD. We used the YCSB framework \cite{ycsbc_bench,ycsb} for experimental evaluation with a dataset size of approx. 50GB, unless stated otherwise. The WT cache size is set to 100MB and LSM-chunks as well as partitions are allowed to grow up to 20MB. Direct IO is enabled and the OS page cache is cleaned every second in order to ensure repeatable, reliable and even conservative results.

\begin{figure}[b]
	\begin{minipage}{.65\linewidth}
		\centering
		\subfloat[Relative space and write amplification]{\label{fig:eval_wa_sa}\adjustbox{trim={.0\width} {.43\height} {.0\width} {.0\height},clip,width=\textwidth}
			{\includegraphics{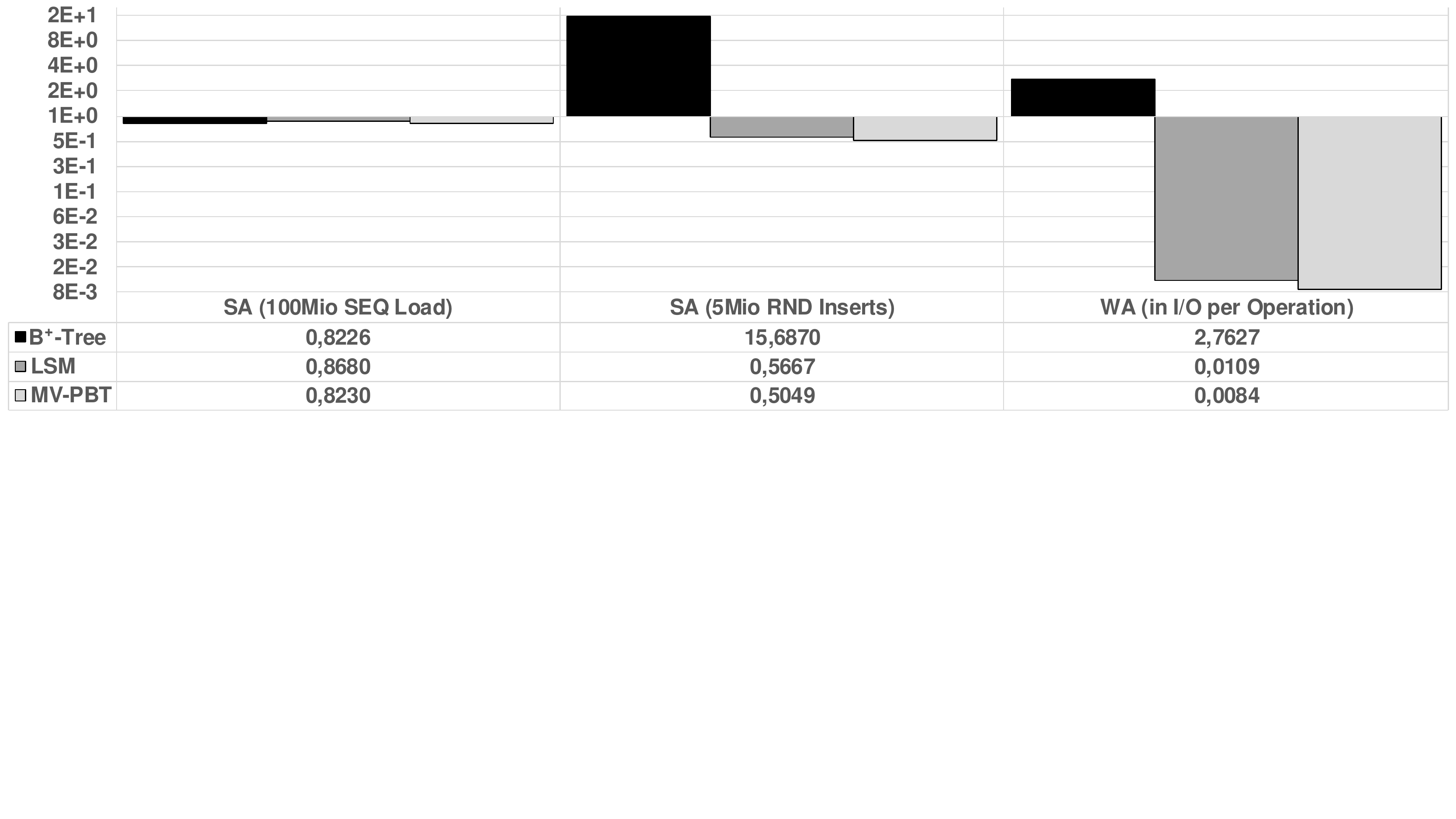}}}
	\end{minipage}
	\begin{minipage}{0.33\linewidth}
		\centering
		\subfloat[MV-PBT tree walk flush]{\label{fig:pbt_seq_wt}\adjustbox{trim={.02\width} {.56\height} {0.02\width} {.0\height},clip,width=\textwidth}{\includegraphics[width=\textwidth]{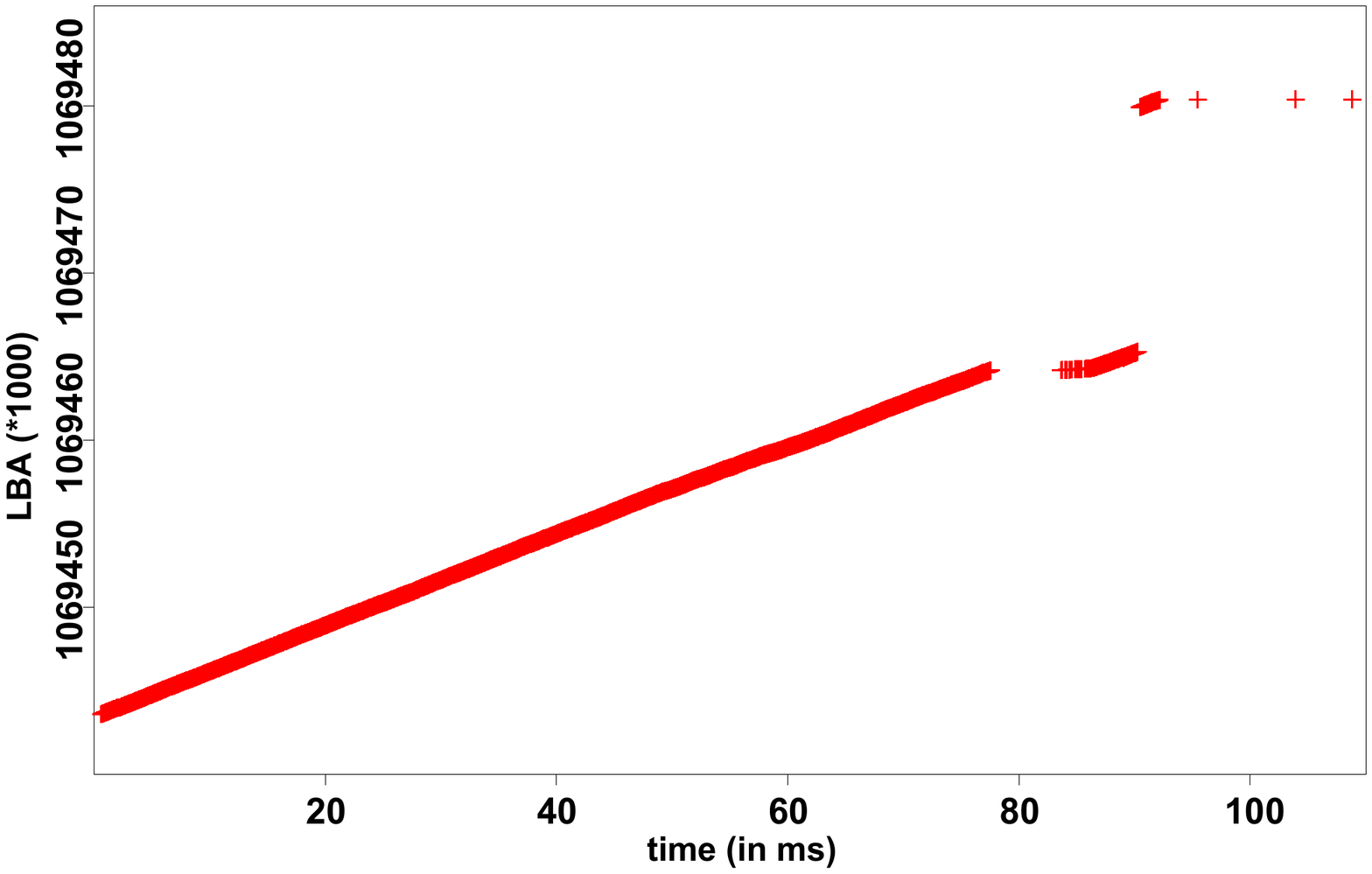}}}
	\end{minipage}
\vspace{-0.25cm}
	\caption{Experiments 1 and 2 evaluate the structural properties of MV-PBT.}
	\label{fig:pbt_wa_sa}
\end{figure}

\subsubsection{Experiment 1: Space and Write Amplification.} In Fig. \ref{fig:eval_wa_sa}, B$^+$-Tree, LSM-Tree (merges are disabled for comparability) and MV-PBT are initially bulk loaded with 100 million records (key and value size are 13 and 16 bytes respectively). Prefix truncation in record keys, suffix truncation in separator keys and snappy compression allow comparable relative space requirements for all approaches. There is a clear evidence of the synergy between prefix truncation and partitioned key, since the enlarged record key by a 2 byte partition number does not result in higher space requirements. Subsequently, 5 million new records are inserted -- yielding in approx. 60 new partitions / LSM-components. Due to compression techniques, the additional relative space requirement is lower than the actually added record size, with slight advantages for MV-PBT. B$^+$-Tree suffer from insertions in the read-optimized layout due to node splits -- yielding in massive relative space amplification per newly inserted records. {\sf Insight:} MV-PBT offers the lowest space amplification, that is between 12\% and 31$\times$ better.
Finally, the write amplification (Fig. \ref{fig:eval_wa_sa}) is evaluated after 5 million inserts. Since almost each insertion causes escalating node splits in the read-optimized layout of a B$^+$-Tree, each insertion causes 2.76 write I/Os of half filled nodes. Sequential writes of dense-packed nodes allow LSM-Trees and MV-PBT to achieve singular writes of optimally filled nodes, yielding in much less write I/O per insertion. MV-PBT achieves a better factor due to commonly used inner nodes. Moreover, merge operations of LSM components would cause a downturn of write amplification by orders of magnitude. {\sf Insight:} compared to LSM-trees, MV-PBT offers 30\% less write amplification and is up to 300$\times$ better than B-Trees.

\subsubsection{Experiment 2: Sequential Write Pattern.} Fig. \ref{fig:pbt_seq_wt} depicts a sequential write pattern with the logical block addresses (LBA) on the ordinate and evolving time on the abscissa. As a result of the partition switch operation, delayed maintenance operations (splits) on leaves followed by inner nodes are performed in a reconciliation operation. Afterwards, leaves are identified by a tree walk and ascending written to secondary storage devices, depicted by the continuously ascending markers. Finally, the referencing levels of immutable inner nodes are sequentially written, depicted by multiple shorter continuously ascending markers. {\sf Insight:} MV-PBT is able to perform advantageous sequential writes.

\begin{figure}[t]
	\begin{minipage}{.595\linewidth}
		\centering
		\subfloat[Accumulated executed transactions (*1k) in YCSB Workload A with 1kB value size]{\label{fig:eval_cum}\adjustbox{trim={.01\width} {.3\height} {.0\width} {.01\height},clip,width=\textwidth}{\includegraphics{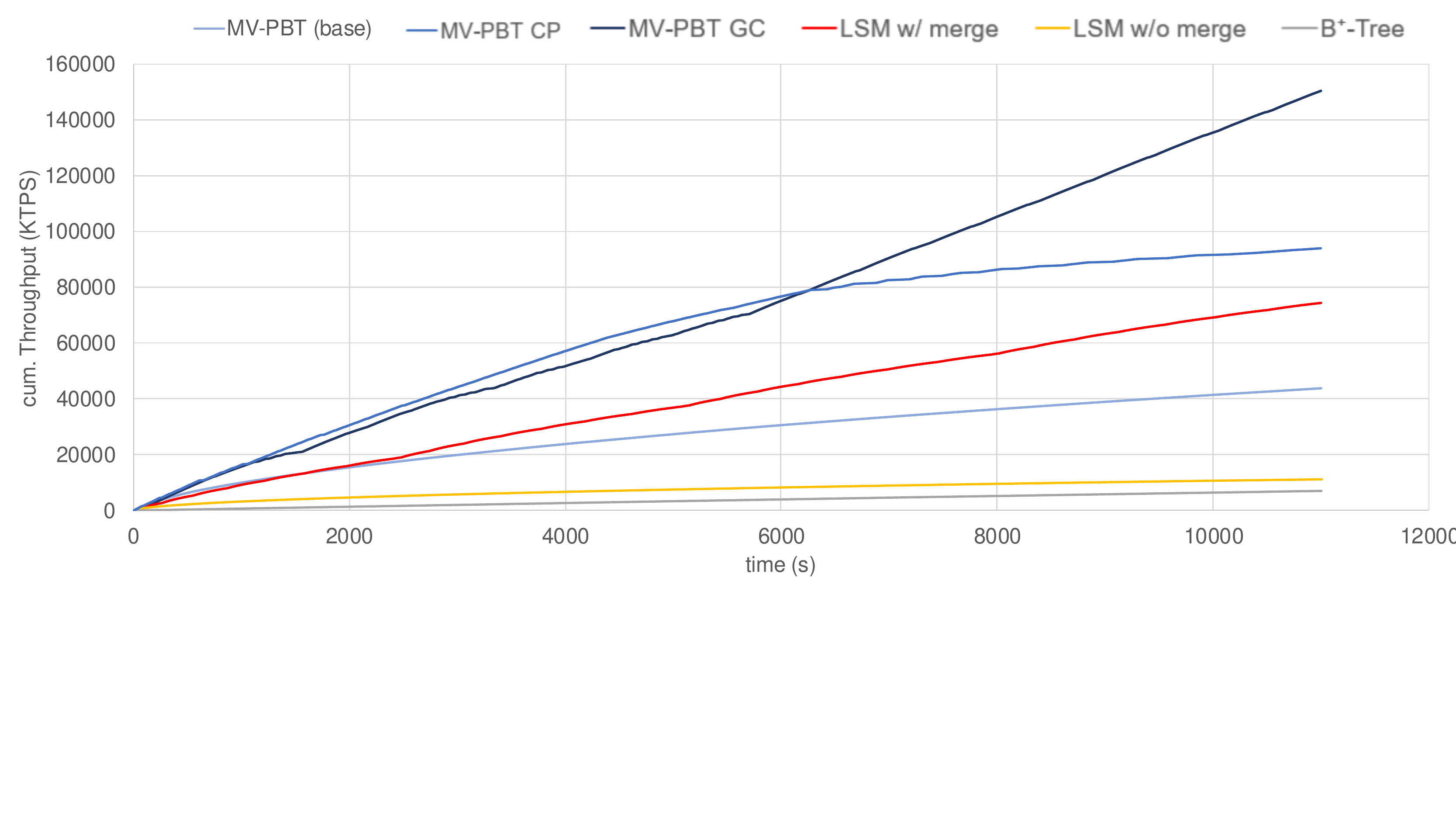}}}
	\end{minipage}
	\begin{minipage}{.395\linewidth}
		\centering
		\subfloat[Intermediate State Read-Only Throughput]{\label{fig:eval_read}\adjustbox{trim={.01\width} {.46\height} {.48\width} {.01\height},clip,width=\textwidth}{\includegraphics{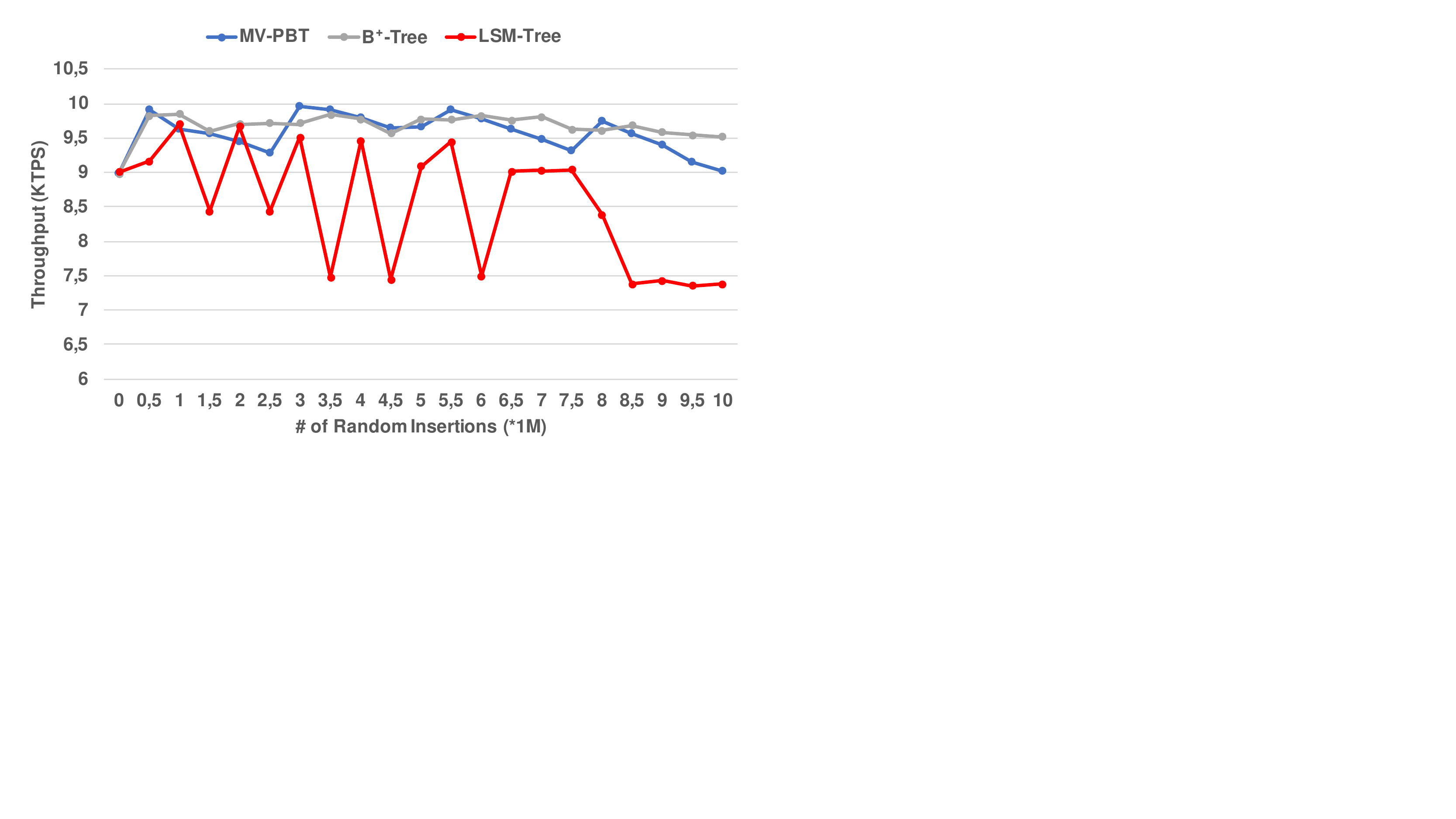}}}		
	\end{minipage}
\vspace{-0.25cm}
\caption{Experiments 3 and 4 evaluate consistent performance of MV-PBT.}
\end{figure}

\subsubsection{Experiment 3: Steady Performance by Cached Partitions and Garbage Collection.} The write-heavy YCSB Workload A consists of 50\% updates and reads, respectively (depicted in Fig. \ref{fig:eval_cum}). Write amplification in B$^+$-Trees yield in poor performance characteristics (7M tx). Sequential writes and low write amplification in base MV-PBT (no Cached Partition and GC) allow much higher transactional throughput, however, increasing search effort degenerates performance (44M tx), whereby LSM-Trees hold search effort down by merges (74M tx). {\sf Insight:} the direct structural comparison of LSM-Trees and MV-PBT is without merges and garbage collection, whereby MV-PBT outperforms LSM (11M tx) by 4$\times$. Enabling Cached Partitions allow MV-PBT increased read efficiency, however, memory footprint of auxiliary filter structures degenerates its capabilities over time due to effectively reduced cache (94M tx). {\sf Insight:} occasional Garbage Collection in MV-PBT (every 400 Partitions) enables stable performance characteristics (151M tx), outperforming LSM-Trees by 2$\times$. 

\subsubsection{Experiment 4: Read-Only Performance Characteristics of intermediate Structures States.} YCSB Workload C is performed several times after inserting 500k small random records for 10 minutes, respectively (depicted in Fig. \ref{fig:eval_read}). B$^+$-Tree remain very stable, but slightly decrease, since the read-optimized layout breaks. LSM-Trees throughput is varying based on the number of LSM components. {\sf Insight:} commonly cached inner nodes and periodically created Cached Partitions allow MV-PBT to retain comparable read performance even if 80 partitions are created after 10 million random insertions.

\begin{figure}[t]
\begin{minipage}{.325\linewidth}
	\centering
	\subfloat[YCSB Workload A]{\label{fig:eval_a}\adjustbox{trim={.01\width} {.58\height} {.65\width} {.01\height},clip,width=\textwidth}{\includegraphics{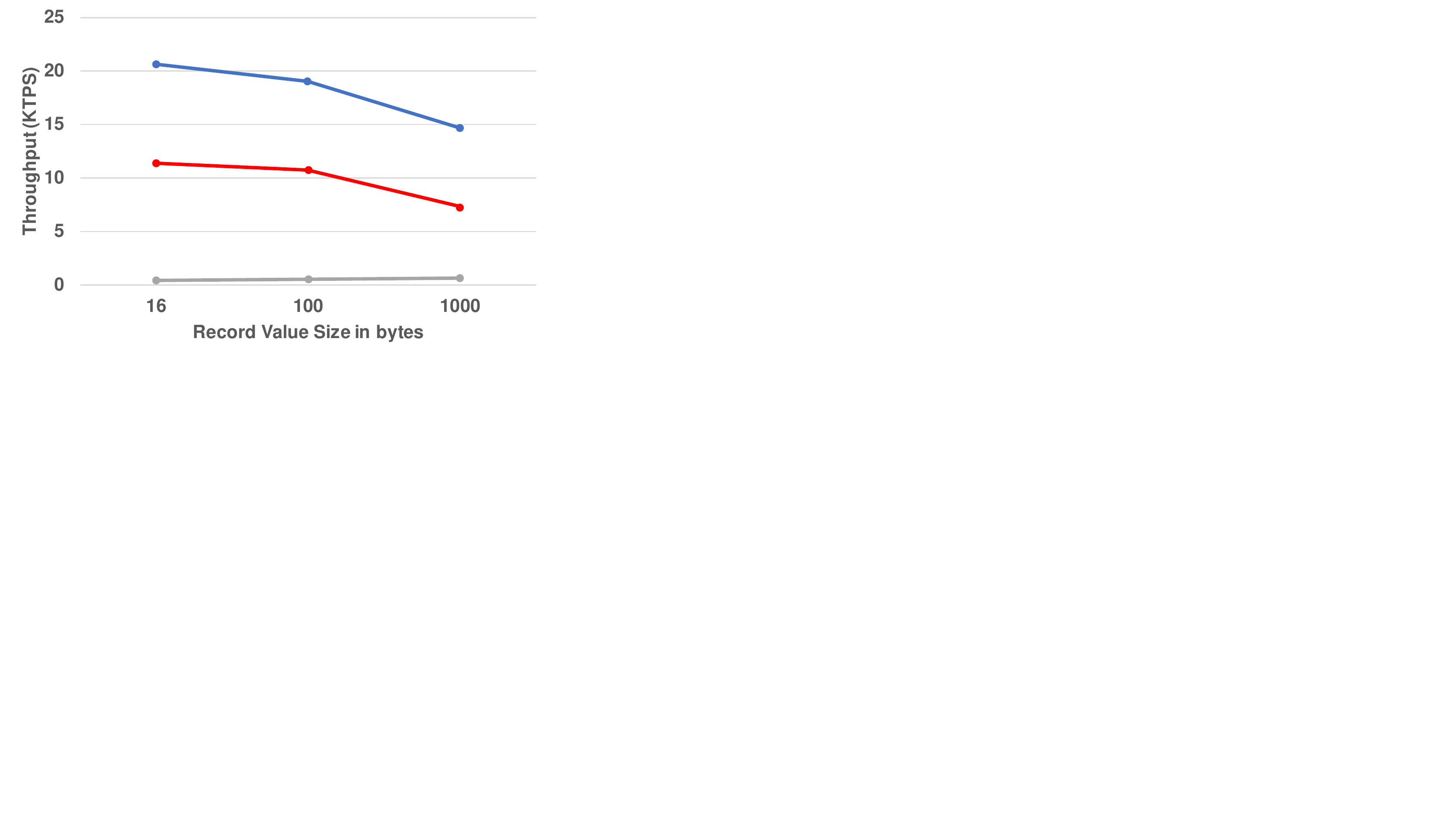}}}
\end{minipage}
	\begin{minipage}{.325\linewidth}
	\centering
	\subfloat[YCSB Workload B]{\label{fig:eval_b}\adjustbox{trim={.01\width} {.58\height} {.65\width} {.01\height},clip,width=\textwidth}{\includegraphics{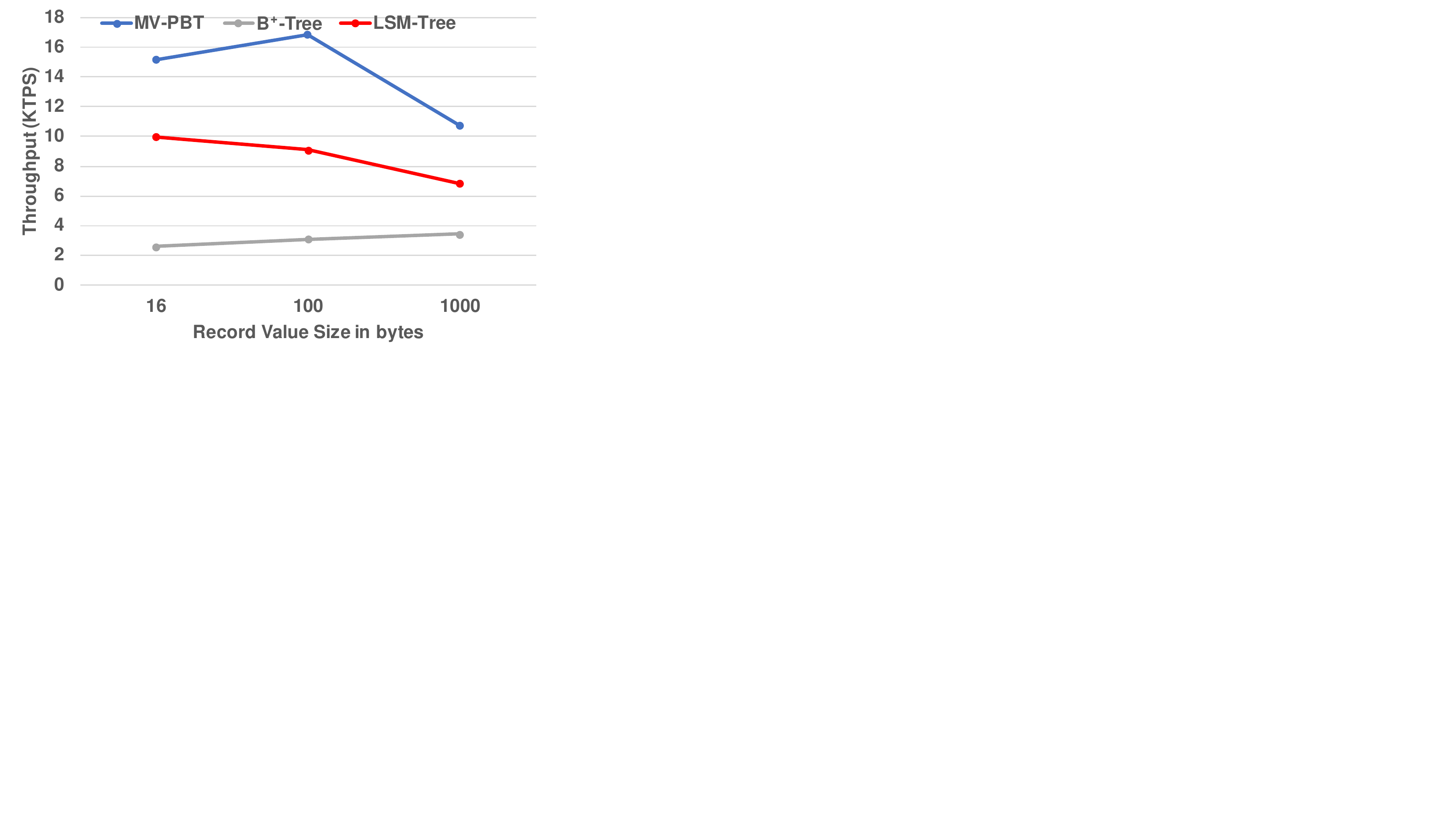}}}
\end{minipage}
	\begin{minipage}{.325\linewidth}
	\centering
	\subfloat[YCSB Workload C]{\label{fig:eval_c}\adjustbox{trim={.01\width} {.58\height} {.65\width} {.01\height},clip,width=\textwidth}{\includegraphics{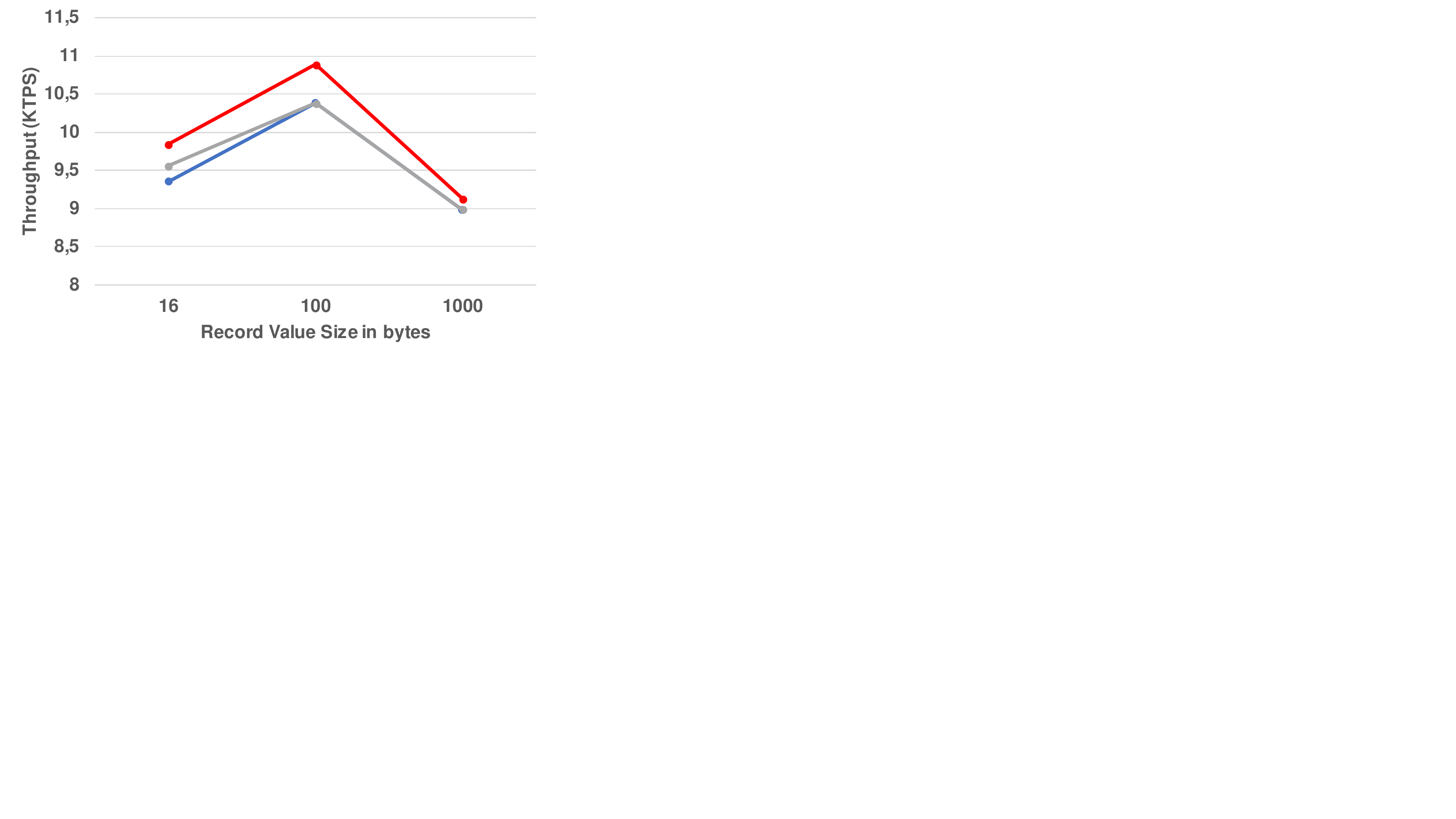}}}
\end{minipage}

\centering
	\begin{minipage}{.325\linewidth}
	\centering
	\subfloat[YCSB Workload D]{\label{fig:eval_d}\adjustbox{trim={.01\width} {.58\height} {.65\width} {.01\height},clip,width=\textwidth}{\includegraphics{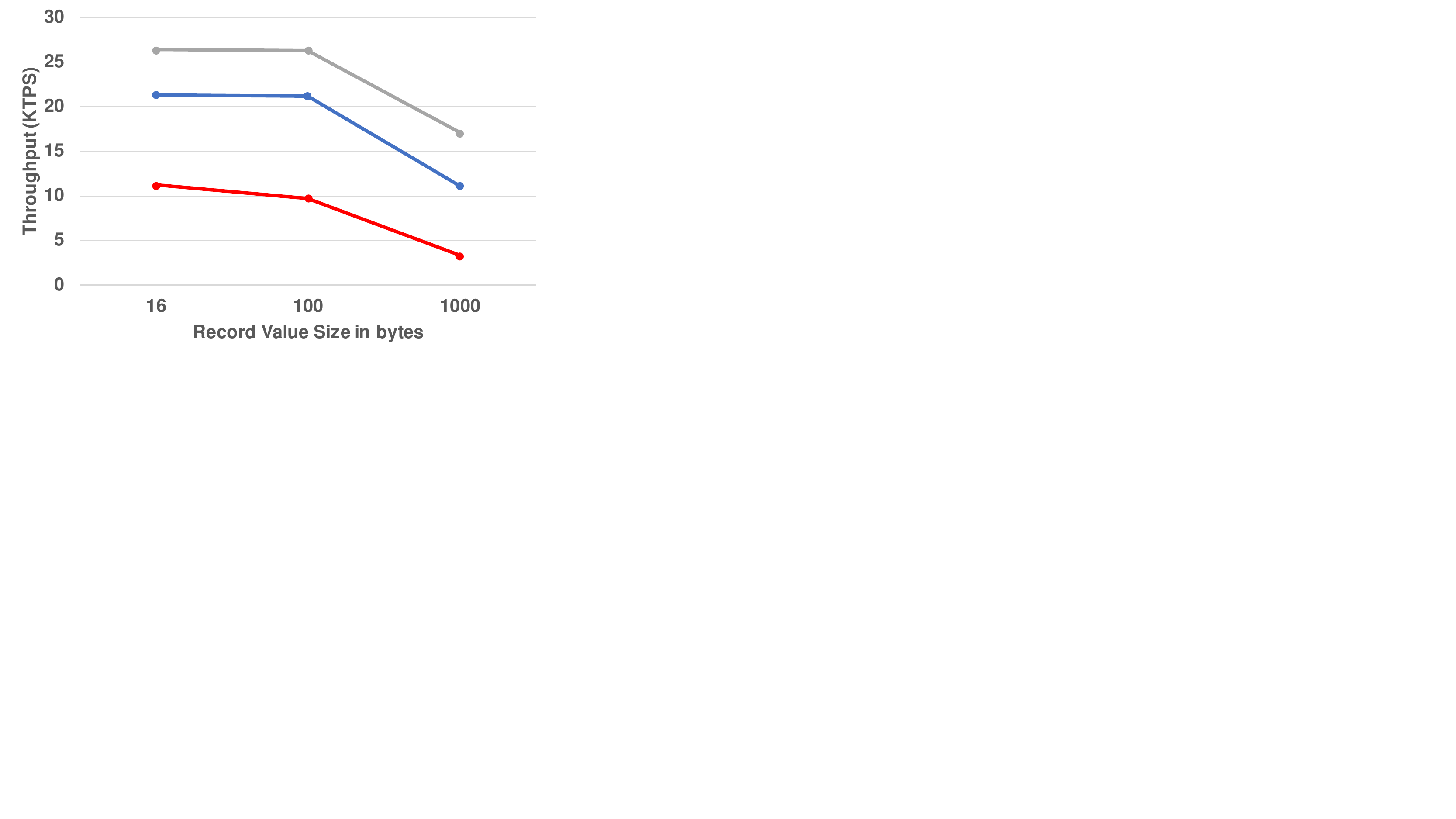}}}
\end{minipage}
	\begin{minipage}{.325\linewidth}
	\centering
	\subfloat[YCSB Workload E]{\label{fig:eval_e}\adjustbox{trim={.01\width} {.58\height} {.65\width} {.01\height},clip,width=\textwidth}{\includegraphics{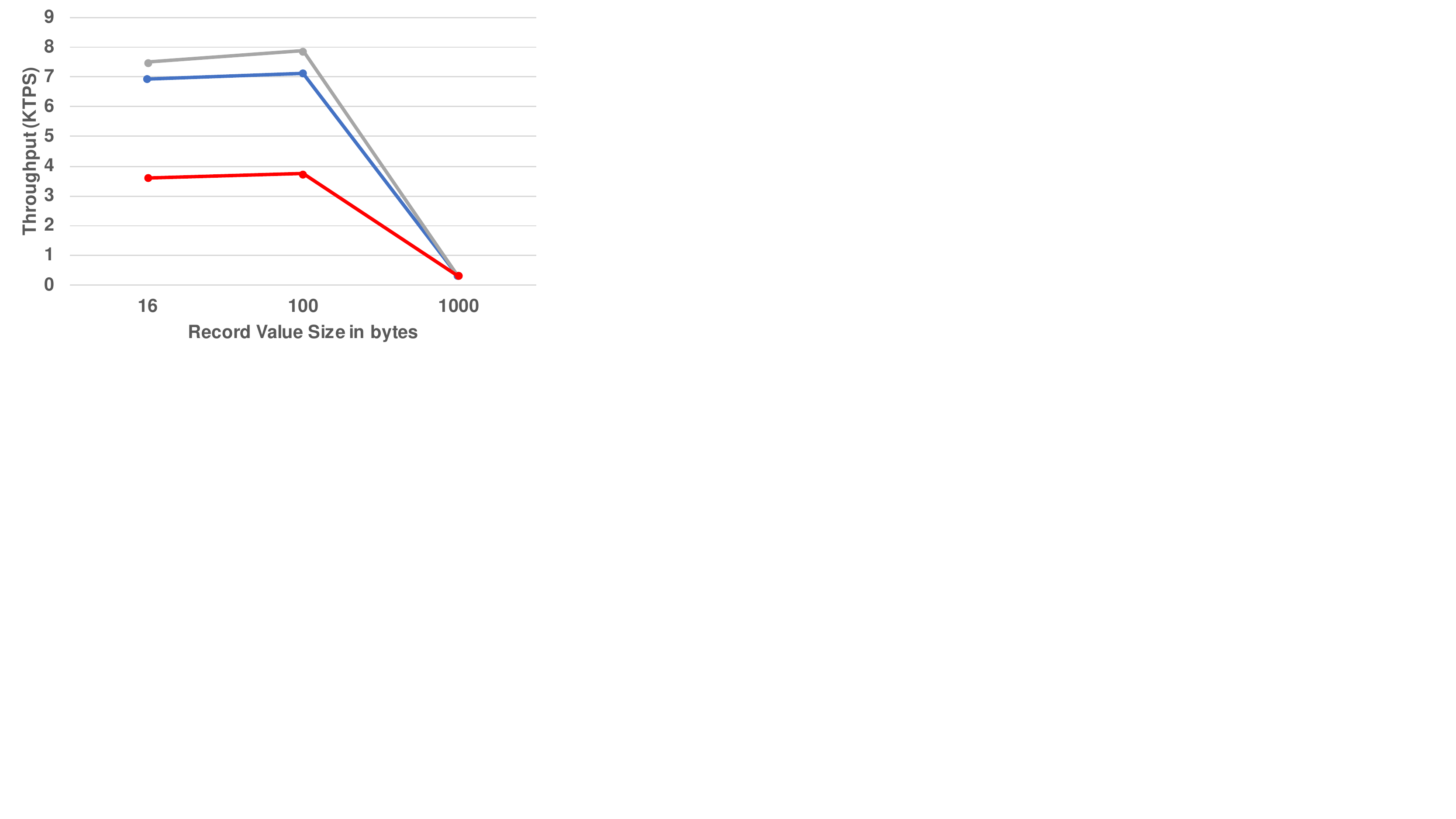}}}
\end{minipage}
\vspace{-0.25cm}
	\caption{Experiment 5 evaluates performance for different value sizes.}
	\label{fig:ycsb}
\end{figure}

\subsubsection{Experiment 5: Impact of Different Value Sizes.} YCSB basic workloads (Fig. \ref{fig:ycsb}) are performed on small (16 bytes), medium (100 bytes) and large (1000 bytes) value sizes, the initial load has been adjusted to match approx. 50GB dataset size. {\sf Insight:} MV-PBT outperforms its competitors in the high and medium update intensive workloads A and B, even the LSM-Tree by 2$\times$ in the workload A. The read-only workload C is performed on the read-optimized layout after load phase -- comparable results prove negligible costs of partitioned key comparisons, whereas LSM-Trees are only able to retain performance for one component (compare Fig. \ref{fig:eval_c} and \ref{fig:eval_read}). Workload D searches for few concurrently inserted records. B$^+$-Tree benefits from well cached nodes in the traversal path due to the recent insertion. This is also valid for MV-PBT and LSM-Trees, however, concurrent insertions are not in the MVCC snapshot and cause search operations in other partitions or components, which is 2$\times$ faster in MV-PBT. Finally, MV-PBT is able to achieve comparable performance to B$^+$-Tree in the mostly scan workload E. Cached Partitions and commonly cached inner nodes enable cheap merge sort scan operations.

\section{Conclusion}
\label{sec:conclude}

In this paper we present Multi-Version Partitioned BTrees (MV-PBT) as a sole storage and index management structure \cite{crizz_htap} in KV-storage engines. Logical horizontal partitioning yields beneficial appends of version records within a single tree structure. Partitions leverage properties of B$^+$-Trees by common utilization and caching of inner nodes in traversal operations, whereby constant search performance and high fragmentation are brought together. This behavior leveraged by Cached Partition in order to minimize write amplification to secondary storage devices. Contrary to LSM-Trees, merging is considered for garbage collection of obsolete version records than for sustained search performance, wherefore MV-PBT is predestinated to be applied in KV-storage engines.

 \bibliographystyle{splncs04}
 \bibliography{pbtvslsm}
\end{document}